\documentclass[aps,pra,twocolumn,showpacs,groupedaddress,eqsecnum]{revtex4}
\usepackage{graphicx}
\usepackage{dcolumn}
\usepackage{bm}

\begin{document}


\title{Quantitative conditional quantum erasure in two-atom resonance 
fluorescence}
\author{Matthias Jakob$^{1,3}$}
\author{J\'anos Bergou$^{1,2}$}
\affiliation{$^{1}$Department of Physics, Hunter College, 
City University of New York, 695 Park Avenue, New York, NY 10021, USA}
\affiliation{$^{2}$Institute of Physics, Janus Pannonius University,
H-7624 P\'{e}cs, Ifj\'{u}s\'{a}g \'{u}tja 6, Hungary}
\affiliation{$^{3}$Department of Physics, Royal Institute of 
Technology (KTH), SCFAB, Roslagstullsbacken 21, S-10691 Stockholm, Sweden}
\date{\today}
\begin{abstract}
We present a conditional quantum eraser which erases the a
priori knowledge or the predictability of the path a photon
takes in a Young-type double-slit experiment with two fluorescent
four-level atoms. This erasure violates a recently derived erasure relation
which must be satisfied for a conventional, unconditional quantum eraser
that aims to find an optimal {\it sorting} of the system into
subensembles with particularly large fringe visibilities. The conditional
quantum eraser employs an interaction-free, partial which-way
measurement which not only sorts the system into
optimal subsystems with large visibility but also selects
the appropriate subsystem with the maximum possible visibility.
We explain how the erasure relation can be violated
under these circumstances. 
\end{abstract}

\maketitle

\section{introduction}\label{sec.1}
The quantum eraser \cite{scully1} has been approved as a remarkable tool
for studying fundamental topics in quantum mechanics. Among them we mention 
complementarity \cite{bohr}, the understanding of measurements 
in quantum mechanics, and, most importantly, the entanglement 
between subsystems which enables the very concept of quantum
erasure \cite{scully2,englert1,abranyos1,jakob}.
There have been several experimental realizations of the quantum eraser
demonstrating complementarity as one of the most basic principles
of quantum mechanics \cite{zajonc,kwiat,zeilinger1,Pittman,Duerr,Kim}.
Further, quantitative measures of complementarity have been derived on
the basis of an inequality for (partial) predictability $\cal{P}$
of the path the particle takes and (partial) visibility $\cal{V}$
of the interference fringes in a two-way interferometer 
\cite{Zurek,Greenberger,Mandel} and experimentally verified \cite{Rauch}.
Somewhat later, Englert \cite{englert} as well as
Jaeger, Shimony, and Vaidman \cite{Jaeger} derived an inequality which 
quantifies complementarity in a two-way interferometer 
supplemented by a which-way detector.
They introduce the new quantity {\textit{distinguishability}} 
$\cal{D}$ which quantifies the maximum possible information, 
obtainable by the which-way detectors, that Nature can grant us about the 
path the particle actually takes.
The inequality has also been verified in several subsequent experiments 
\cite{Duerr1,Tsegaye}.
It has been recognized that the which-path knowledge ${\cal{K}}$
which is the practically available which-way information
granted by the which-way detectors may be considerably less than the
distinguishability ${\cal{D}}$ in realistic which-path measurements
\cite{bjoerk,abranyos2,englert2}.

The inequalities have been extended in order
to incorporate quantum erasure by introducing the quantum eraser
visibility, ${\cal{V}}^{\left(\text{QE}\right)}$ \cite{englert2}, 
or the equivalent conditioned visibility, ${\cal{V}}_{c}$ \cite{bjoerk}.
The results can be summarized by the erasure relation \cite{englert2}
\begin{equation}
{\cal{P}}^{2}+{\cal{C}}^{2} \leq 1, \label{eq1.1}
\end{equation}
where ${\cal{C}}$ is the coherence which denotes the maximum possible
quantum eraser visibility, ${\cal{V}}^{\left(\text{QE}\right)}\leq {\cal{C}}$,
Nature can grant us about the visibility of the interference fringes.
Taking into account that the smallest value of the actual which-path
information, ${\cal{K}}$, is given by the a priori knowledge,
${\cal{P}}$, the following inequality inequality has been derived
\cite{englert2},
\begin{equation}
{\cal{K}}^{2}+({\cal{V}}^{\left(\text{QE}\right)})^{2} \leq 1, \label{eq1.2}
\end{equation}
which states that the quantum eraser visibility 
${\cal{V}}^{\left(\text{QE}\right)}$ can not exceed the 
maximum value $\left(1-{\cal{P}}^2\right)^{1/2}$,
since ${\cal{P}}\leq {\cal{K}}$.
Thus, a conventional quantum eraser can not erase the a priori predictability 
${\cal{P}}$ which fundamentally limits the performance of quantum erasure.

In this paper we demonstrate erasure of the a priori 
predictability with a {\it conditional} quantum eraser. 
The retrieved visibility explicitly exceeds the maximum possible value 
derived for conventional quantum erasers [see Eq.\ (\ref{eq1.1})]. 
The conditional quantum eraser not only does sort subensembles as
conventional quantum erasers do but, beyond it, also selects a proper
subensemble. This can be achieved by employing interaction-free {\it
partial} measurements, as has been suggested in the paper by Elitzur
and Dolev \cite{Elitzur} who consider
nonlocal effects of partial measurements and quantum erasure.
We extend their results to erasure of a priori knowledge and derive
the underlying physical concept which enables predictability to be
erased and which turns out to be {\textit{concurrence}} or 
{\textit{two-particle visibility}}. 
The concept of concurrence and especially two-particle visibility 
yields an interesting relation between entanglement and complementarity. 
Further, the quality of the recovered interference 
fringes will strongly depend on the amount of concurrence in the initial 
system. The partial interaction-free measurement composes
a non-unitary transformation which, evidently, 
has a certain probability of failure. 
However, we can always and unambiguously tell whether 
the transformation (erasure) has succeeded.

\begin{figure}[h,t]
\includegraphics[width=7cm]{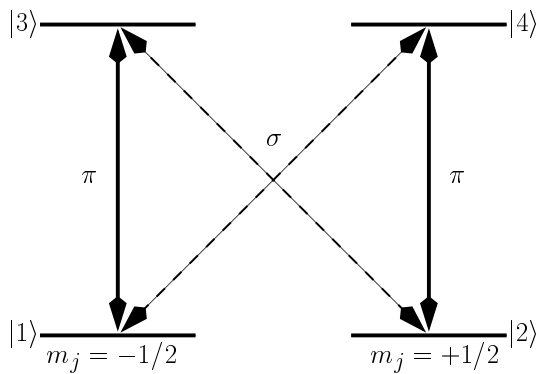}
\caption{Internal structure of the four-level atom with
relevant polarization-sensitive transitions. 
Transitions between the states
$|1\rangle\leftrightarrow|3\rangle$ and
$|2\rangle\leftrightarrow|4\rangle$ which preserve 
the internal magnetic quantum number $m_{j}$ are connected with 
$\pi$-or $z$-polarized light, while transitions between 
$|1\rangle\leftrightarrow|4\rangle$ and
$|2\rangle\leftrightarrow|3\rangle$ change the magnetic 
quantum number and lead to $\sigma$- or $x$- and $y$-polarized 
scattering events.}
\label{Fig. 1}
\end{figure}
We implement the conditional quantum eraser in an
interference experiment with light scattered from two trapped
four-level atoms. The internal structure of the four-level atom together with
polarization-sensitive transitions is displayed in Fig.\ \ref{Fig. 1}
(see also Ref.\ \cite{schuurman}).
Recently, this remarkable Young-type double-slit experiment 
succeeded in the observation of polarization-dependent 
interference effects \cite{eichmann1} and stimulated ongoing research
\cite{interference}.
We model the partial interaction-free measurement with the help of
inefficient detectors which can be simulated by perfect 
detectors together with beamsplitters.
A detector click indicates a failure of the conditional quantum eraser while
the absence of a click means that the quantum erasure (of predictability)
has succeeded. The success rate will be limited by the degree of
predictability. The quantum eraser visibility will further depend on
the amount of concurrence in the initial system which limits the quality 
of the retrieved interference fringes. 

The structure of the paper is as follows. 
In Section \ref{sec.2} we present the
interferometric system and consider the general idea of conditional
quantum erasing. 
We also discuss differences between the conditional quantum eraser
and conventional quantum erasers. 
Inequalities are derived and compared for the conditional and
conventional quantum eraser. 
We restrict our considerations to pure states of the entangled
system in this section.
In Section \ref{sec.3} we apply the general idea of conditional
quantum erasing to the four-level atom interferometer and discuss
various realizations of conditional quantum erasure in this system. We
derive the success probability
of erasure and discuss its relation to the concurrence in the initial
(unsorted) system.
We extend our results to cases when the initial system
is not in a pure (entangled) state in Section \ref{sec.4} 
and show the limitation on the performance of conditional
quantum erasure due to the mixture. 
In particular, it will turn out that we can not reach
the maximum visibility anymore. Finally, in Section \ref{sec.5}, we
present some concluding remarks and discussions.
%
%
%
%
%
\section{Conditional quantum erasure: general consideration}\label{sec.2}
\begin{figure}[h,t]
\includegraphics[width=8.6cm]{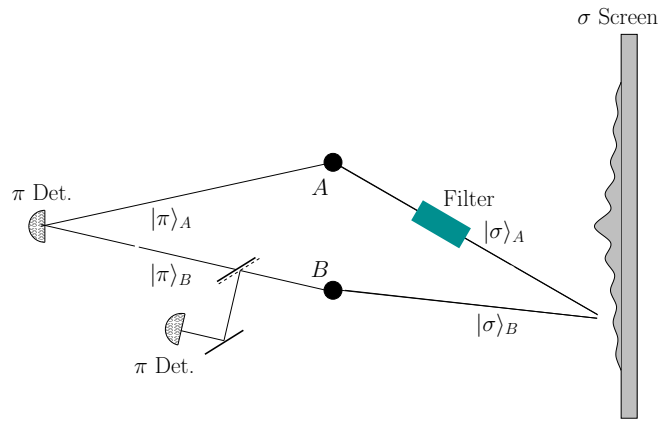}
\caption{Interferometric set-up of the conditional quantum eraser. 
An optical filter, with a transmittance $t$, in the
upper right arm of the interferometer generates (partial) predictability 
about the path the interfering $\sigma$-photon takes. 
The measurement device in the lower left arm, consisting of a
beamsplitter with transmittance $t_{\text{BS}}$ and a 
photon detector, performs a partial, interaction-free which-path 
measurement on the erasing $\pi$-photon.  
If the detector clicks, the conditional quantum eraser failed. 
If the detector does not respond, the conditional quantum 
eraser succeeded. In this case the correlated measurement 
of the erasing photon at an equidistantly positioned  
detector in between the two atoms and the interfering photon will result 
in a complete erasing of the predictability under certain circumstances. }
\label{Fig. 2}
\end{figure}

We first discuss and summarize the general idea of 
the conditional quantum eraser on the basis of a simple model 
in an attempt to prepare the reader for the arguments which follow.
Suppose we are given two spatially separated two-photon sources,
$A$ and $B$ ( see Fig.\ \ref{Fig. 2}), that generate the following
polarization-entangled photon state
\begin{equation}
|\Psi\rangle_{A,B}=|\pi\rangle_{A}\otimes|\sigma\rangle_{A}+
|\pi\rangle_{B}\otimes|\sigma\rangle_{B}, \label{eq2.1}
\end{equation}
where $\pi$ and $\sigma$ denotes different polarization
and the indexes $A$ and $B$ denote the origin of the photons.
We stress already here that the entanglement between the photons is an 
essential ingredient for the nonclassical features of the conditional
quantum eraser. 
The coupled system (\ref{eq2.1}) contains $\sigma$-polarized photons,
forming the ``interfering system'', which are detected at an
interference screen, 
and the $\pi$-polarized photons which establish the
``environmental degrees of freedom'' and whose detection may serve as, e.g.\
a which-path measurement.
In the ``environment'' only those degrees of freedom 
are included which can be controlled
by the experimentalist by, e.g., which-path measurements 
or quantum eraser sorting.

Let us first discuss the $\sigma$-polarized part 
of the system on the basis of the interference experiment 
displayed in Fig.\ \ref{Fig. 2}.
Suppose, we only consider the right part of the interference configuration in
Fig.\ \ref{Fig. 2} which establishes a standard Young-type 
interference experiment with two atoms acting as the double-slit. 
First, we assume just one $\sigma$-polarized photon emitted by the
two atoms and not the entangled photon system of Eq.\ (\ref{eq2.1}).
In addition, one of the two paths (the upper one in Fig.\ \ref{Fig. 2})
is partially blocked by an optical filter. The intensity of the light 
or, to be more precise, 
the probability amplitude of the photon will be reduced, 
compared to the other optical path (the lower one in Fig.\ \ref{Fig. 2}),
as
\begin{equation}
|\sigma\rangle=\frac{1}{\sqrt{1+t}}\Big(
\sqrt{t}|\sigma\rangle_{A}+|\sigma\rangle_{B}\Big). \label{eq2.2}
\end{equation}
Here $t$ is the transmittance of the optical filter.
The reduction obviously depends on the transmittance
$t$. Consequently, partial, {\it a priori} knowledge about
the path the photon actually takes is granted and this is expressed
by the {\it predictability} ${\cal{P}}$
\cite{Zurek,Greenberger,Mandel,Rauch,englert,Jaeger,Duerr1,Tsegaye,bjoerk,abranyos2,englert2}
\begin{equation}
{\cal{P}}=\left|\frac{1-t}{1+t}\right|=\frac{1-t}{1+t}. \label{eq2.3}
\end{equation}
We expect the visibility ${\cal{V}}$ to be reduced as a consequence of
this a priori knowledge in agreement with the general duality relation
\begin{equation}
{\cal{P}}^2+{\cal{V}}^2\leq 1, \label{eq2.4}
\end{equation}
leading to the following maximum possible visibility in this interferometer
\begin{equation}
{\cal{V}}_{\text{max}}=\sqrt{1-\left(\frac{1-t}{1+t}\right)^{2}}
=\frac{2}{1+t}\sqrt{t}. \label{eq2.5}
\end{equation}

Let us now consider the complete interferometer including
the environmental, $\pi$-polarized photon but neglecting the
partial measurement device in the lower left path. The entanglement
between the photons in Eq.\ (\ref{eq2.1}) enables us to perform
which-path measurement 
or quantum sorting on the environment, i.e.\ on the $\pi$-photon and
this, in turn, alters the interference properties of the interference
photon. In particular, as a consequence of the entanglement, 
we will not observe any interference fringes of the 
$\sigma$-photons if we do not detect the $\pi$-photon
\begin{eqnarray}
\langle E_{A,\sigma}^{\dagger}E_{B,\sigma}+A\leftrightarrow B\rangle
&=& \langle \pi_{A}|\pi_{B}\rangle\langle
\sigma_{A}|E_{A,\sigma}^{\dagger}E_{B,\sigma}|\sigma_{B}\rangle
\nonumber \\
&&+A\leftrightarrow B,
\nonumber \\
&=& 0. \label{eq2.5a}
\end{eqnarray}
The entanglement between the subsystems enables us to get {\it possible}
which-path information expressed by the quantity {\it
distinguishability}, ${\cal{D}}$, which limits interference
visibility according to the duality relation
\cite{englert,Jaeger}
\begin{equation}
{\cal{D}}^{2}+{\cal{V}}^{2}\leq 1. \label{eq2.6}
\end{equation}
The distinguishability is the maximum {\it possible}
which-path information Nature can grant us and it is, 
in this interferometer, given by ${\cal{D}}=1$. 
This explains why we do not observe interference when we neglect
the environment, i.e.\ do not detect or measure the $\pi$-photon.
The actual knowledge, ${\cal{K}}$, which is the amount of which-path
information that we can learn from the measurement of the environment,
is usually smaller
than the distinguishability, ${\cal{D}}$ \cite{englert2}. In
particular, when we 
do not measure the $\pi$-photon, the actual which-way information ${\cal{K}}$
is just given by the a priori knowledge or predictability ${\cal{P}}$,
Eq.\ (\ref{eq2.3}), since we do not gain further
information if we discount the environment. 
In general, the actual which-path information is limited by the following
inequality \cite{englert2}
\begin{equation}
{\cal{P}}\leq{\cal{K}}\leq{\cal{D}}. \label{eq2.7}
\end{equation}

One of the reasons, why we start with the entangled state, (\ref{eq2.1}), is
that quantum eraser sorting in the usual sense is immediately possible.
This can be done, if we detect the environmental $\pi$-photon with a detector
equidistantly placed from the two atoms as indicated in Fig.\ \ref{Fig. 2}.
A correlated detection of the $\sigma$- and $\pi$-photons 
erases the distinguishability and we retrieve interference. 
However, the recovered {\textit{quantum eraser visibility}},
${\cal{V}}^{\left(\text{QE}\right)}$, of the interference fringes is limited
if the a priori knowledge ${\cal{P}}$ is not equal to zero.
The recovered quantum eraser visibility for the interferometer 
in Fig.\ \ref{Fig. 2} is given by
\begin{equation}
{\cal{V}}^{\left(\text{QE}\right)}=\frac{2}{1+t}\sqrt{t}, \label{eq2.8}
\end{equation}
where $t$ is, again, the transmittance of the optical filter and where
we took into account the following entangled state of the system 
under consideration
\begin{equation}
|\Psi\rangle_{A,B}=\frac{1}{\sqrt{1+t}}\left(\sqrt{t}
|\pi\rangle_{A}\otimes|\sigma\rangle_{A}+
|\pi\rangle_{B}\otimes|\sigma\rangle_{B}\right) . \label{eq2.8a}
\end{equation}
In particular, the quantum eraser visibility is limited 
by the {\it coherence} which is, in the case of the pure entangled 
state under consideration, just the concurrence
${\cal{C}}=\frac{2}{1+t}\sqrt{t}$ [see Eq.\ (\ref{eq2.8a})]
\cite{wootters}, and
\begin{equation}
{\cal{V}}\leq{\cal{V}}^{\left(\text{QE}\right)}\leq{\cal{C}}. \label {eq2.9}
\end{equation}
Here ${\cal{V}}$ denotes the ``a priori visibility'' of the system 
before quantum erasing \cite{englert2}.
We note at this point, that the quantity ${\cal{C}}$ was also called the
coherence in Ref. \cite{englert2}. 
Although a possible relation of this quantity
to an entanglement measure was mentioned, it was not 
associated with the {\it concurrence} introduced by Wootters
\cite{wootters}. 
Here, we close this gap by stating that the ``coherence'' is
indeed a measure of entanglement which is explicitly given 
by the concurrence of the initial system.
We also mention that the concurrence is directly connected with 
the two-particle visibility introduced by Jaeger {\it et al.} 
\cite{Jaeger} (see also \cite{sergienko}) which gives the concurrence 
an evident physical meaning.  
 
Englert and Bergou demonstrated the following {\it erasure relation} 
\cite{englert2}, on the basis of the inequalities (\ref{eq2.7}) and 
(\ref{eq2.9}),
\begin{equation}
{\cal{P}}^{2}+{\cal{C}}^{2}\leq 1. \label{eq2.10}
\end{equation}
Thus, the maximum possible visibility of the recovered interference
fringes, 
${\cal{V}}^{\left(\text{QE}\right)}_{\text{max}}$, can never reach 
unity if the predictability, ${\cal{P}}$, is not equal to zero,
\begin{equation}
{\cal{V}}^{\left(\text{QE}\right)}_{\text{max}}\leq \sqrt{1-{\cal{P}}^{2}}.
\label{eq2.11}
\end{equation}
In other words, a conventional quantum eraser
can never erase the {\it a priori} predictability ${\cal{P}}$.
When we consult Eqs.\ (\ref{eq2.3}) and (\ref{eq2.8}) it is easy to
see that the considered interferometric
scheme satisfies the general inequalities, (\ref{eq2.10}) and (\ref{eq2.11}). 
Moreover, since the state under consideration is a pure state, 
the inequalities transform into equalities \cite{englert2}.

We have just showed that a conventional quantum eraser can not erase
the a priori knowledge or predictability, ${\cal{P}}$. 
Does this mean that it is impossible to erase predictability? 
We anticipate the answer by claiming that it is
indeed possible to erase predictability. 
However, this can only be achieved with a certain success 
probability which is subject to some constraints.
We, therefore, call such an eraser a {\it conditional quantum eraser}.
In the following, our task will be to find a device
which erases the a priori knowledge.
A naive approach starts, again, by considering the right part
of the interferometric device in Fig.\ \ref{Fig. 2}. 
We assume just a single $\sigma$-polarized photon which arises  
from the sources $A$ and $B$ and do not consider a biphoton at first. 
This approach will also help us to understand the
crucial quantum features of the conditional {\textit{quantum}} eraser. 
Suppose we utilize a second optical filter with the same 
transmittance, $t$, as the upper optical filter 
in the lower optical path of the right part
of the interferometric device in Fig.\ \ref{Fig. 2}. 
Accordingly, the initial state of the photon, (\ref{eq2.2}), is
transferred to 
\begin{equation}
|\sigma\rangle =\frac{1}{\sqrt{2t}} 
\left(\sqrt{t}|\sigma\rangle_{A} + \sqrt{t}|\sigma\rangle_{B}\right). 
\label{eq2.14}  
\end{equation} 
Thus, we recover perfect visibility of the interference fringes and 
apparently erase the a priori knowledge.  
However, at the same time we change the a priori conditions of the 
interferometric setup. The second optical filter alters the
interference device to an ideal two-way interferometer with  
no a priori knowledge about the path the photon takes. 
We have gained perfect fringe visibility by changing 
the interferometric device for the interfering $\sigma$-photon.  
In other words, we alter the a priori condition of the right part of the
interferometer which defines the initial interferometric properties of the 
$\sigma$-photon.
This, however, does not constitute the basis what we want to consider as 
an erasing of the a priori knowledge. 
In particular, we do not want to change the right part of the 
interferometric setup which alters the a priori conditions of the 
interfering $\sigma$-photon and consequently changes the a priori 
knowledge about the path the photon takes.  

The concept to erase predictability without directly or locally 
affecting the interfering photon, in the sense of changing the 
a priori conditions of the right interferometer in 
Fig.\ \ref{Fig. 2}, is realized by the quantum property
{\textit{entanglement}}.  
Thus, we have to consider a biphoton as, e.g.\ that of Eq.\ (\ref{eq2.1}), 
for the arguments which follow. 
As already mentioned, entanglement between 
the photons is essential in order to meet the requirements for the
conditional quantum eraser.   
Again, a naive approach starts with changing the a priori properties 
of the complete four-port interferometer in Fig.\ \ref{Fig. 2}.
That is, we insert an optical filter in the lower left arm
of the interferometer with the same transmittance $t$ as 
the one in the upper right arm. 
Although classically this will not have any 
effect on the interfering $\sigma$-photon, quantum mechanically, due 
to the entanglement in the biphoton state (\ref{eq2.1}), the biphoton 
is transformed into the following state
\begin{equation}
|\Psi\rangle_{A,B} \to \frac{1}{\sqrt{2t}} 
\left( \sqrt{t} |\pi\rangle_{A}\otimes|\sigma\rangle_{A} + 
\sqrt{t} |\pi\rangle_{B}\otimes|\sigma\rangle_{B} \right). 
\label{eq2.15}
\end{equation} 
Consequently, we recover full interference fringes in a
coincidence measurement of the two photons, i.e.\ under the quantum 
eraser conditions. However, the initial properties of the four-port 
interferometer have changed albeit under an  
intriguing, {\textit{nonlocal}} and certainly nonclassical manner. 
The predictability of the path of the {\textit{biphoton}} has become
zero in Eq.\ (\ref{eq2.15}) due to an optical intervention. 

This is still not completely the situation which we want to consider
as a conditional quantum eraser. We try to realize the conditional
quantum eraser without changing the initial optical properties of the
four-port interferometer.  This can be achieved with the help of a
nonlocal, partial and interaction-free which-way measurement of the
path the photon takes \cite{Elitzur}. As in the previous
discussion, we assume a black box, representing the measurement
device, instead of the optical filter in the lower left 
arm of the four-port interferometer. The measurement device will not
locally  
interact with the interfering photon but only affects the ''erasing
photon''.  
Thus, in the same manner as above, the entanglement between the
interfering  
and the erasing photon plays an essential role in the manipulation of
the optical properties of the interfering photon by the erasing photon.
We will, at this point, not discuss details about the black box or
measurement device but only argue about its general features. 
We assume a measurement device that can partly determine the path of  
the erasing photon ($\pi$-photon) in the four-port interferometer. 
In addition, the measurement apparatus can report a positive 
or a negative result. That is, the measurement apparatus can
unambiguously tell if it succeeded in partially distinguishing between
the paths the $\pi$-photon took (positive result) or not (negative
result). Further, the probability of failure is adjustable to the value 
predetermined by the predictability of the path of the 
interfering $\sigma$-photon, $1-t/(1+t)$. 
If we denote the states of the measurement device with
$|0\rangle_{\text{M}}$,  
related with a positive result or success, and $|1\rangle_{\text{M}}$, 
connected with a negative result or failure, 
the state of the biphoton including the states of the measurement 
apparatus unitarily transforms into 
\begin{eqnarray}
|\Psi\rangle
&\to&
\sqrt{\frac{1}{1+t}}
\left[
\left(\sqrt{t}|\sigma\rangle_{A}\otimes|\pi\rangle_{A}\right.\right. 
\nonumber \\
&&\left. + \sqrt{t}|\sigma\rangle_{B}\otimes|\pi\rangle_{B}\right)
\otimes|0\rangle_{\text{M}} \nonumber \\
&&  + 
\Big(\sqrt{1-t}|\sigma\rangle_{B}\otimes|\pi\rangle_{B}\Big)
\otimes|1\rangle_{\text{M}}\Big]. \label{eq2.16}
\end{eqnarray}
This constitutes the initial state of the conditional quantum eraser. 
Suppose the measurement apparatus gives a positive result. Under this 
condition the state of the biphoton is transformed in such a way that 
we recover perfect fringe visibility for the $\sigma$-photon 
under quantum erasing conditions, i.e.\ in a coincidence measurement 
of the biphoton. The partial, interaction-free which-way measurement 
of the $\pi$-photon manipulates the entangled $\sigma$-photon 
in such a manner that full interference fringes are 
recovered in a quantum eraser scheme.

The partial interaction-free measurement becomes evident 
if we consider the effect of the measurement on the $\pi$-photon itself. 
Clearly, the initial state of the $\pi$-photon, 
$|\pi\rangle_{A}+|\pi\rangle_{B}$ is transformed into 
$|\pi\rangle_{A}+\sqrt{t}|\pi\rangle_{B}$ under the influence of the 
partial, interaction-free measurement. 
If, on the other hand, the measurement apparatus responds (negative
result) we will discard the biphoton.   
In other words, we have erased the predictability 
or the a priori knowledge of the path the $\sigma$-photon takes 
with the help of an interaction-free, partial which-way measurement
of the $\pi$-photon. 
However, this can only be achieved under the condition 
that the measurement apparatus does not react. 
The probability that this takes place is connected with the initial 
predictability of the $\sigma$-photon path. In particular the failure 
probability of our conditional quantum eraser is directly given 
by the predictability, ${\cal{P}}$.
The realization of the measurement apparatus or the black box will be the 
subject of the next section. We will also make clear why we can talk
about an interaction-free measurement. 
%
%
\section{Conditional quantum erasure in a two-atom interferometer}\label{sec.3}
We start with the consideration of the two-atom, four-port interferometer 
in Fig.\ \ref{Fig. 2}. 
Suppose the initial atomic state is given as
\begin{equation}
|\Psi\rangle_{\text{atom}}=|1\rangle_{A}\otimes|1\rangle_{B}. 
\label{eq3.1}
\end{equation}
Applying a $\pi$-polarized laser pulse with weak enough intensity that 
only one atom at any given time will be excited, the atoms transform into 
\begin{equation}
|\Psi\rangle_{\text{atom}}\to |1\rangle_{A}\otimes|3\rangle_{B}+
|3\rangle_{A}\otimes|1\rangle_{B}. \label{eq3.2}
\end{equation}
Let us assume that the detector on the right in our four-port interference 
scheme is only sensitive to $\sigma$-polarized photons, i.e.\ we ignore
$\pi$-polarized photons. 
In this case the system evolves into the following state  
\begin{eqnarray}
|\Psi\rangle_{\text{atom}}
&\to& 
|\Psi\rangle_{\text{atom}}\otimes|\Psi\rangle_{\text{ph}}, \nonumber \\
&=&
|1\rangle_{A}\otimes|2\rangle_{B}\otimes|\sigma\rangle_{B}+
|2\rangle_{A}\otimes|1\rangle_{B}\otimes|\sigma\rangle_{A}, \nonumber \\
&& \label{eq3.3}
\end{eqnarray}
after decay of the atoms.  
Next, we apply a second, $\sigma^{+}$ circularly polarized laser pulse, again 
weak enough that only one atom at a time will be excited.
This transforms the atomic state into
$|\Psi\rangle_{\text{atom}}=|1\rangle_{A}\otimes|3\rangle_{B}+
|3\rangle_{A}\otimes|1\rangle_{B}$.  
Now, we assume the detector in the left part of our four-port interferometer 
only to be sensitive to $\pi$-polarized photons. That is, we ignore  
$\sigma$-polarized photons in the second transition.
Under this condition, after the decay of the atoms, 
the system will evolve into the following state 
\begin{eqnarray}
|\Psi\rangle_{\text{atom}}\otimes|\Psi\rangle_{\text{ph}}
&\to&
|1\rangle_{A}\otimes|1\rangle_{B}\otimes
|\sigma\rangle_{B}\otimes|\pi\rangle_{B} 
\nonumber \\
&&+
|1\rangle_{A}\otimes|1\rangle_{B}\otimes
|\sigma\rangle_{A}\otimes|\pi\rangle_{A}. 
\nonumber \\
&& \label{eq3.4}
\end{eqnarray}
We ignored the time-dependence of the photon arrivals as well 
as the time-dependence of the application of the 
laser pulses throughout the above derivation for simplicity reasons. 
Further, the time-dependence of the photon arrivals is not 
important for the considerations which follow. 
We notice, that the two-atom interferometer can generate an
entangled state between the photons identical to the entangled photon
state in Eq.\ (\ref{eq2.1}). 
From now on, we neglect the atomic subspace in (\ref{eq3.4}) which will 
not change after generation of the biphoton state and 
reaches the initially assumed atomic state after each generation of
biphotons, ready for another generation cycle. 
 
Let us first neglect the measurement device in the lower left arm of the
four-port interferometer in Fig.\ \ref{Fig. 2}. The optical filter in
the upper  
right arm of the interferometer generates a priori knowledge of the 
$\sigma$-polarized photon in dependence of its transmittance $t$.
In particular, the $\sigma$-photon state by itself will be transformed 
into $|\sigma\rangle_{A}+|\sigma\rangle_{B}\to
(\sqrt{t}|\sigma\rangle_{A}+|\sigma\rangle_{B})/\sqrt{(1+t)}$ which 
clearly shows that the partial predictability of the path 
the photon takes is given by ${\cal{P}}=1-t/(1+t)$. 
Translating this to the biphoton state (\ref{eq3.4}) and ignoring the 
atomic subspace we arrive at the expression, 
\begin{eqnarray}
|\Psi\rangle_{\text{ph}}
&\to&
\sqrt{\frac{1}{1+t}}
\Big(
|\sigma\rangle_{B}\otimes|\pi\rangle_{B}
+\sqrt{t}
|\sigma\rangle_{A}\otimes|\pi\rangle_{A}\Big). \nonumber \\
&& \label{eq3.5}
\end{eqnarray}
In other words, in a coincidence measurement of the biphoton with the 
detector of the $\pi$-photon equidistantly placed from the two atoms we
realize the {\textit{conventional}} quantum eraser condition
\cite{abranyos1}. 
The ability to recover interference fringes is limited by the
a priori knowledge about the path the $\sigma$-photon takes. 

In order to implement the {\textit{conditional}} quantum eraser we 
have to insert the measurement device. 
In particular, we want to realize a partial, interaction-free 
which-way measurement of the $\pi$-photon. 
A possible device capable in doing this is 
already implied in the lower left arm of the four-port interferometer,
in Fig.\ \ref{Fig. 2}. 
It consists of a beamsplitter which splits the path of the 
$|\pi\rangle_{B}$-photon into two alternatives. 
We place a detector, which operates with 
100$\%$ efficiency, in one of the two alternatives.
Suppose the ratio between transmission and reflection of the beamsplitter 
can be adjusted to a preset value. In particular, we can set 
the transmittance of the beamsplitter, $t_{\text{BS}}$,
equal to the transmittance of the optical filter, $t_{\text{BS}}=t$,
in the upper right arm of the interferometer.

Let us consider how this affects the $\pi$-photon, i.e.\ we ignore for
a moment the $\sigma$-photon and consider the $\pi$-photon alone.
Due to the unitary transformation of the beamsplitter, the
$\pi$-photon transforms into the following state
\begin{eqnarray}
\frac{1}{\sqrt{2}}
\Big(|\pi\rangle_{A}+|\pi\rangle_{B}\Big) &\to&
\frac{1}{\sqrt{2}}\Big[\Big(
|\pi\rangle_{A}+\sqrt{t_{\text{BS}}}|\pi\rangle_{B}^{(1)}\Big)\otimes
|0\rangle_{\text{M}}
\nonumber \\
&&\qquad\;\;
+\sqrt{1-t_{\text{BS}}}|\pi\rangle_{B}^{(2)}\otimes
|1\rangle_{\text{M}}\Big]. 
\nonumber \\
&& \label{eq3.6}
\end{eqnarray}
Here we took into account the detector states $|0\rangle_{\text{M}}$, 
indicating success, and $|1\rangle_{\text{M}}$
for failure (the detector clicks), as well as the two alternatives 
$|\pi\rangle_{B}^{(1)}$ and $|\pi\rangle_{B}^{(2)}$ of the
$|\pi\rangle_{B}$-photon amplitude. The alternative 
$|\pi\rangle_{B}^{(1)}$, of course, is the original mode or 
the original ''path'' the $\pi$-photon takes. 
It is evident from Eq.\ (\ref{eq3.6}) that the beamsplitter 
together with the detector performs a partial and interaction-free 
which-way measurement of the path the $\pi$-photon takes in 
the interferometer. 
In particular, when no interaction with the detector takes place, 
i.e.\ the detector is in the $|0\rangle_{\text{M}}$ state which 
is the outcome corresponding to success, we gain partial information
about the $\pi$-photon path. The obtained knowledge is given by 
${\cal{K}}=(1-t_{\text{BS}})/(1+t_{\text{BS}})$. 
On the other hand, the partial measurement fails if the detector 
reacts, i.e.\ a photon is detected, indicated by the detector state
$|1\rangle_{\text{M}}$. In other words, a partial, interaction-free 
measurement of the $\pi$-photon takes place under the condition that the 
detector does not click. Consequently, we are allowed to speak
about a {\textit{conditional}}, interaction-free
and partial which-way measurement. 

Next, we consider the effect of the partial which-way 
measurement on the biphoton given by Eq.\ (\ref{eq3.5}). 
The biphoton transfers into the following state as a consequence of the 
$\pi$-photon measurement
\begin{eqnarray}
|\Psi\rangle_{\text{ph}}
&\to&
\sqrt{\frac{1}{1+t}}
\Big[\Big(\sqrt{t_{\text{BS}}}|\sigma\rangle_{B}\otimes|\pi\rangle_{B}
+\sqrt{t}|\sigma\rangle_{A}\otimes|\pi\rangle_{A}\Big)
\nonumber \\
&&
\otimes|0\rangle_{\text{M}}
\nonumber \\
&&
+\sqrt{1-t_{\text{BS}}}|\sigma\rangle_{B}\otimes|\pi\rangle_{B}^{(2)}\otimes
|1\rangle_{\text{M}}
\Big], 
\nonumber \\
&=& 
\sqrt{\frac{1}{1+t}}
\Big[\sqrt{t+t_{\text{BS}}}
\Big(\sqrt{\frac{{t_{\text{BS}}}}{{t+t_{\text{BS}}}}}
|\sigma\rangle_{B}\otimes|\pi\rangle_{B}
\nonumber \\
&&
+\sqrt{\frac{{t}}{{t+t_{\text{BS}}}}}
|\sigma\rangle_{A}\otimes|\pi\rangle_{A}\Big)
\otimes|0\rangle_{\text{M}}
\nonumber \\
&&
+\sqrt{1-t_{\text{BS}}}|\sigma\rangle_{B}\otimes|\pi\rangle_{B}^{(2)}\otimes
|1\rangle_{\text{M}}
\Big].
\label{eq3.7}
\end{eqnarray}
Here, we did not explicitly label alternative one of the $\pi$-photon
since it is the original path, introduced previously. From Eq.\
(\ref{eq3.7}) we see that it is possible to recover perfect fringe 
visibility of the $\sigma$-photon in a coincidence measurement with the 
$\pi$-photon detected at an equal distance from the two atoms. 
Thus, the ``no-click'' event in the measurement device 
indicates a ``click'' event on the eraser detector. This fact is
important in practice if inefficient detectors are involved. 
In case of ideal conditions, we detect all involved photons 
and the ``no-click'' event on the measurement device is 
equivalent with a ``click'' event on the eraser detector.  
However, this can only be achieved when, first, the transmittance of 
the beamsplitter is equal to the transmittance of 
the optical filter, $t_{\text{BS}}=t$, and, second, under the condition 
that the photon detector of the measurement apparatus does 
not respond, i.e.\ the measurement device works 
interaction free and resides in the state $|0\rangle_{\text{M}}$. 
In other words we have implemented the {\textit{conditional quantum eraser}} 
with the help of the partial, interaction-free measurement device.
We have erased the predictability of the path the $\sigma$-photon takes. 
We further verify, that the failure probability of the partial measurement 
device is determined by the a priori predictability ${\cal{P}}=1-t/(1+t)$ 
of the path the $\sigma$-photon takes in case of 
optimal erasing conditions, i.e.\ when $t_{\text{BS}}=t$.
Moreover, the product of success probability and two-particle visibility
or concurrence of the obtained quantum state, (\ref{eq3.7}), is limited 
by the degree of concurrence or two-particle visibility
of the initial state, (\ref{eq2.8a}) (see also discussion following Eq.\ 
(\ref{eq2.8a})),
\begin{equation}
\frac{{t+t_{\text{BS}}}}{1+t}\frac{2\sqrt{tt_{\text{BS}}}}{t+t_{\text{BS}}}
=\frac{2\sqrt{tt_{\text{BS}}}}{1+t} 
\leq {\cal{C}}=\frac{2}{1+t}\sqrt{t}, \label{eq3.8}
\end{equation}
since $0\leq t_{\text{BS}}\leq1$. In Eq.\ (\ref{eq3.8}) the two-particle 
visibility or concurrence is given by 
${\cal{C}}_{\text{cond}}=\frac{2\sqrt{tt_{\text{BS}}}}{t+t_{\text{BS}}}$,
and  
the probability of success is given by $S=\frac{{t+t_{\text{BS}}}}{1+t}$. 
We mention, that the two-particle visibility of the obtained quantum
state, (\ref{eq3.7}), is identical to the ''single-particle
visibility'' in 
the conditional quantum eraser, ${\cal{V}}^{(\text{QE})}_{\text{cond}}
={\cal{C}}_{\text{cond}}$ of the $\sigma$-photon in case of quantum 
erasing condition, i.e.\ in a coincidence measurement of the biphoton 
with a $\pi$-photon detector equidistantly placed from the
two atoms $A$ and $B$. 
We mention also, that the conditional quantum eraser visibility, 
${\cal{V}}^{(\text{QE})}_{\text{cond}}$, strongly depends on the
transmittance 
$t_{\text{BS}}$. When the conditional quantum eraser 
does not work optimally, i.e.\ $t_{\text{BS}}\neq t$, the conditional 
quantum eraser visibility will be smaller than one,
\begin{equation}
{\cal{V}}^{(\text{QE})}_{\text{cond}}=
\frac{2\sqrt{tt_{\text{BS}}}}{t+t_{\text{BS}}}
< 1, \label{eq3.9}
\end{equation}
and when the conditional quantum eraser 
works optimally, i.e.\ for $t_{\text{BS}}=t$, the conditional quantum 
eraser visibility reaches one, 
\begin{equation}
{\cal{V}}^{(\text{QE})}_{\text{cond}}=
\frac{2\sqrt{tt_{\text{BS}}}}{t+t_{\text{BS}}}
= 1. \label{eq3.10}
\end{equation}
Equations\ (\ref{eq3.9}) and \ref{eq3.10}), again,
clearly indicate that the conditional quantum eraser visibility 
may exceed the quantum eraser visibility of a conventional quantum eraser. 
In particular, the conditional quantum eraser visibility can
explicitly top the concurrence or two-particle visibility 
of the {\textit{initial}} system which is impossible with a 
conventional quantum eraser [see Eq.\ (\ref{eq2.9})]. 

The complementarity relation (\ref{eq2.10}) represents 
the main feature of the conditional quantum eraser. 
It contains all the important facets of the physical peculiarities. 
First, we get a direct relation between entanglement and complementarity. 
In particular, the concurrence is a well known entanglement measure 
\cite{wootters} and it is, in the case of the 
four-port two-atom interferometer, given by the measurable 
quantity of two-particle visibility \cite{Jaeger}.
A similar expression for real particles has also been found in 
\cite{bose}, but only with the equal sign. 
In addition, the expression there contains distinguishability
and not predictability, which arises because of 
the entanglement of the biparticle to an additional subsystem.
In Eq.\ (\ref{eq2.10}) the concurrence is 
related to a single particle property of one of the two biphotons, i.e.\ 
to the system itself in the form of predictability.
Here, we will, however, not discuss the general properties 
and the general validity of this complementarity relation 
as an entanglement measure, but only some of its relevant
characteristics with relation to the conditional quantum 
eraser. We will discuss a more general version of the 
complementarity relation with regard to entanglement measures and
nonlocality or violations of Bell-type inequalities in another publication. 

We want to verify that the general complementarity relation, Eq.\
(\ref{eq2.10}), 
is fulfilled in the conditional quantum eraser. Although we can 
erase the a priori predictability of the $\sigma$-photon path, this 
does not mean that we can violate the general complementarity relation 
(\ref{eq2.10}). The conditional quantum eraser can violate
the quantum eraser relation for conventional quantum erasers, Eq.\ 
(\ref{eq2.11}), but not the general complementarity relation.
The conditional quantum eraser can change the initial conditions, 
${\cal{P}}$ and ${\cal{C}}$, of the biphoton in a certain subspace 
of the complete biphoton plus detector Hilbert space determined 
by the null result of the detector. 
The ability to achieve this rearrangement in a conditional 
quantum eraser goes beyond the abilities of the conventional quantum
eraser. The predictability, ${\cal{P}}_{\text{cond}}$, and
concurrence, 
${\cal{C}}_{\text{cond}}$, in the conditional quantum eraser, on
condition  
that the detector is in the $|0\rangle_{\text{M}}$ state, are given by
\begin{eqnarray}
{\cal{P}}_{\text{cond}}
&=& \frac{|t-t_{\text{BS}}|}{t+t_{\text{BS}}}, 
\label{eq3.11} \\
{\cal{C}}_{\text{cond}}
&=& 
\frac{2\sqrt{t_{\text{BS}}t}}{t+t_{\text{BS}}}. 
\label{eq3.12}
\end{eqnarray}
Clearly, the two complementary quantities, (\ref{eq3.11}) and
(\ref{eq3.12}), 
fulfill the general complementarity relation, (\ref{eq2.11}), for pure
states 
\begin{eqnarray}
{\cal{P}}_{\text{cond}}^{2} + 
{\cal{C}}_{\text{cond}}^{2} &=&
\frac{(t-t_{\text{BS}})^2}{( t+t_{\text{BS}})^2} +
\frac{4t_{\text{BS}}t}{(t+t_{\text{BS}})^2} , 
\nonumber \\
&=&
1. \label{eq3.13} 
\end{eqnarray}
However, when $t=t_{\text{BS}}$ we can fully erase the
predictability and recover perfect fringe visibility 
clearly indicating the ability of the conditional quantum eraser 
to change the a priori conditions which limit the conventional quantum
eraser.  
In the next section we consider effects of mixture on the conditional 
quantum eraser. We will expose additional limitations of the 
conditional quantum eraser which will be related to the concurrence of the 
initial system.  

\begin{figure}[h,t]
\includegraphics[width=8.6cm]{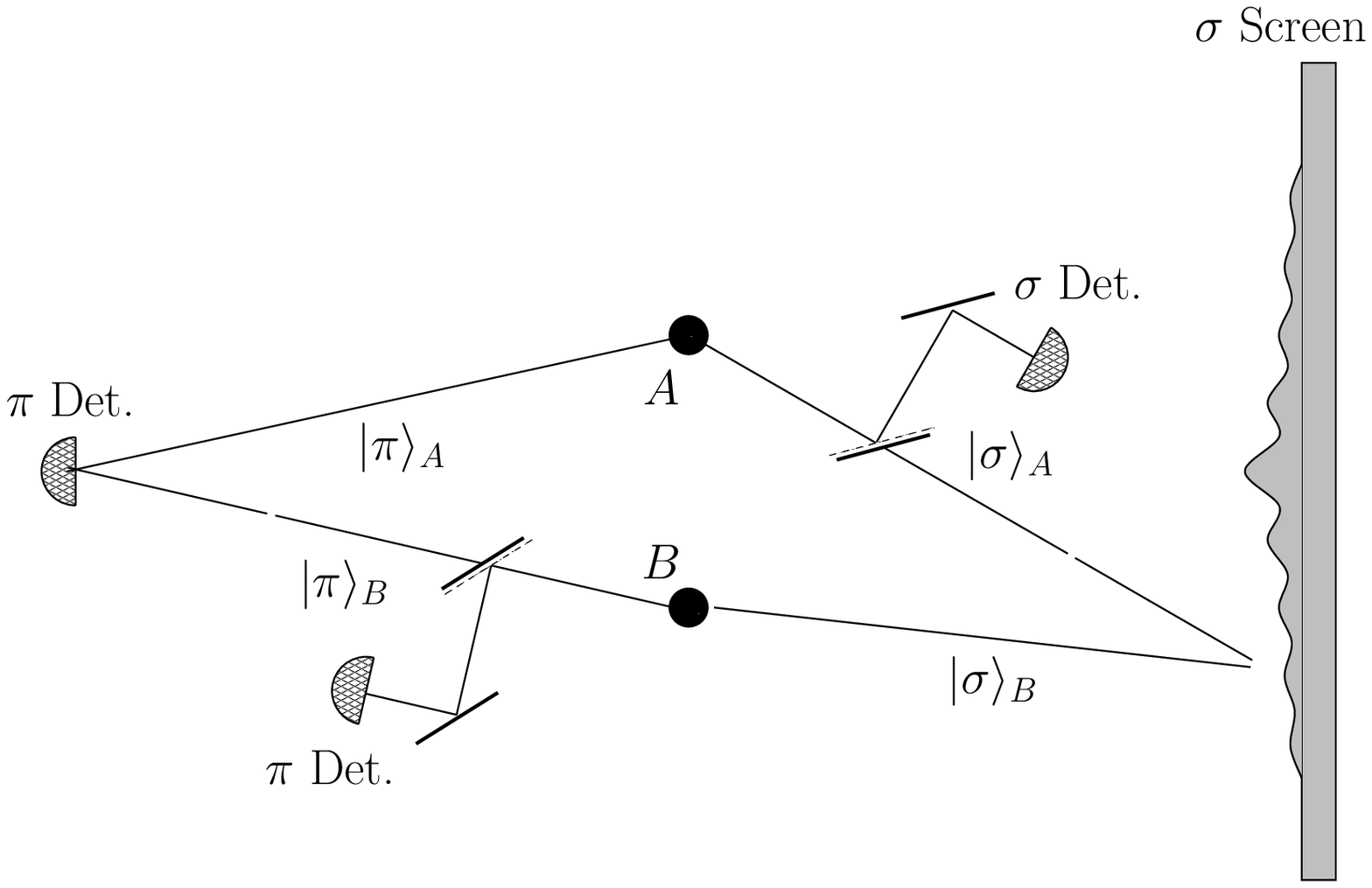}
\caption{Compared to the set-up in Fig.\ \ref{Fig. 2}, this   
interferometer contains an additional partial interaction free 
which-way measurement in the upper right arm. It consists of a beamsplitter  
with transmittance $t_{1}$ and a $\sigma$-detector. Under the 
condition that the detector does not click, we gain  
which-path knowledge about the $\sigma$-photon path.  
This gained knowledge can be similarly erased  
as in Fig. \ref{Fig. 2}. The additional partial which-path  
measurement device consists of a beamsplitter with transmittance 
$t_{2}$ and  a $\pi$-photon detector.}
\label{Fig. 3}
\end{figure}
First, we want to discuss a different conditional quantum
eraser scheme, shown in Fig. \ref{Fig. 3}. It combines two partial 
interaction-free measurements. The first one provides actual knowledge 
about the path the $\sigma$-photon takes, the second one erases the 
obtained which-path information. 
Here, we do not erase the a priori predictability
about the path the $\sigma$-photon takes but the 
obtained knowledge from a conditional, interaction-free measurement. 
In a conventional quantum eraser, the actual which-path information
sets a limit 
on the quantum eraser visibility. We can not erase the actual which-path 
knowledge, ${\cal{K}}$, obtained by the conditional, 
interaction-free measurement with the 
conventional quantum eraser and the maximum possible quantum eraser 
visibility, ${\cal{V}}_{\text{max}}^{\text{QE}}$, is given by 
\begin{equation}
({\cal{V}}_{\text{max}}^{\text{QE}})^{2}\leq 1-{\cal{K}}^{2}. \label{eq3.14}
\end{equation}
In a conditional quantum eraser, however, we can erase the actual
which-path  
knowledge with the help of a second, interaction-free measurement on 
the $\pi$-photons.
Let us consider the effect of the first device which performs
a partial which-way measurement. Suppose the beamsplitter 
has a transmittance of $t_{1}$. This affects the biphoton according to
\begin{eqnarray}
|\Psi\rangle_{A,B} &=& \frac{1}{\sqrt{2}}\left[ \left(\sqrt{t_{1}}
|\sigma\rangle_{A}|\pi\rangle_{A}+|\sigma\rangle_{B}|\pi\rangle_{B}\right)
\otimes|0\rangle_{M1}\right. \nonumber \\
&&
+
\sqrt{1-t_{1}}(|\sigma\rangle_{A}^{\left(2\right)}|\pi\rangle_{A})
\otimes|1\rangle_{M1}\big], \label{eq3.15}
\end{eqnarray}
where we take into account the detector states of the measurement
device and denote the alternative way of the $\sigma_{A}$-photon with 
$|\sigma\rangle_{A}^{\left(2\right)}$. This photon, of course, 
will be absorbed when the detector clicks.  
When the detector does not click, an interaction-free, partial which-way 
measurement of the $\sigma$-photon is accomplished. 
Depending on the transmittance, $t_{1}$, we gain some knowledge
about the path the $\sigma$-photon takes, given by 
\begin{equation}
{\cal{K}}=\frac{1-t_{1}}{1+t_{1}}. 
\label{eq3.16}
\end{equation}
The probability, that the detector does not click, i.e.\ the partial
which-way  
measurement succeeds, is given by $(1+t_{1})/2$. The visibility in a
conventional 
quantum eraser scheme, which consists of the correlated detection of the 
$\pi$- and $\sigma$-photons with a $\pi$-detector equidistantly positioned 
in between the two atoms under the condition that the partial which-path
measurement succeeds, is given by 
\begin{equation}
{\cal{V}}_{\text{max}}^{\text{QE}}=\frac{2\sqrt{t_{1}}}{1+t_{1}}. 
\label{eq3.17}
\end{equation}
Clearly, the conventional quantum eraser can not erase the 
which-path knowledge obtained by the partial which-way measurement. 
The retrieved quantum eraser visibility is limited by the obtained 
knowledge from the partial which-way measurement and the relation
(\ref{eq3.14}) is satisfied
\begin{equation}
({\cal{V}}_{\text{max}}^{\text{QE}})^{2}=\frac{4{t_{1}}}{(1+{t_{1}})^{2}}
=1-{\cal{K}}^{2}=1-\left(\frac{1-t_{1}}{1+t_{1}}\right)^{2}. 
\label{eq3.18}
\end{equation}
On the other hand, the {\textit{conditional}} quantum eraser which 
additionally performs an interaction-free, partial which-way measurement
on the $\pi$-photons can erase the obtained knowledge. This, however, 
can only be achieved with a certain success probability which is limited 
by the obtained which-path knowledge from the first measurement.
Assuming a transmittance of $t_{2}$ of the beamsplitter, the second,
interaction-free partial which-way measurement on 
the $\pi$-photon transforms the state (\ref{eq3.15}) into 
\begin{eqnarray}
|\Psi\rangle_{A,B} &=& \frac{1}{\sqrt{2}}\Big[ \Big(\sqrt{t_{1}}
|\sigma\rangle_{A}|\pi\rangle_{A}+\sqrt{t_{2}}
|\sigma\rangle_{B}|\pi\rangle_{B}\Big)
\nonumber \\
&&
\otimes|0\rangle_{M1}\otimes|0\rangle_{M2} \nonumber \\
&&
+\left(\sqrt{1-t_{2}}
|\sigma\rangle_{B}|\pi\rangle_{B}^{\left(2\right)}\right)
\otimes|1\rangle_{M2} \nonumber \\
&&
+
\left(\sqrt{1-t_{1}}|\sigma\rangle_{A}^{\left(2\right)}|\pi\rangle_{A}\right)
\otimes|1\rangle_{M1}\Big].
\label{eq3.19}
\end{eqnarray}
The conditional quantum eraser can fully recover interference 
fringes with visibility one if the 
transmittances of the beamsplitters are identical, $t_{1}=t_{2}$. 
This, however, only happens with a probability 
of $(t_{1}+t_{2})/2$ in which case none of the detectors clicks. 
Clearly, the success probability goes to zero if the knowledge about the 
path the $\sigma$-photon takes reaches one, i.e.\ the transmittance
$t_{1}\to 0$. 
Similarly to the previous scenario, the product of the success
probability and 
the retrieved interference visibility is limited by the concurrence 
of the biphoton system after the first interaction-free which-way
measurement, which is given by
\begin{equation}
{\cal{C}}=\sqrt{t_{1}}. \label{eq3.20}
\end{equation}
The conditional quantum eraser can erase the partial 
which-way information and fully recover interference fringes as long
as the biphoton system after the first which-path measurement contains 
some concurrence, i.e.\ $\sqrt{t_{1}}>0$. 
It is the (additional) non-unitary evolution initiated
by the partial which-way measurement 
which enables us to recover the full interference fringes in a
conditional quantum eraser. This (additional) non-unitary evolution 
is the reason why the conditional quantum 
eraser outperforms the conventional or traditional quantum eraser. 
We note that the correlated measurement of the $\pi$- and
$\sigma$-photons 
represents also a nonunitary evolution which both the conventional
and conditional quantum eraser have in common. It is this 
additional nonunitary evolution, induced by the partial which-way
measurement in the conditional quantum eraser, that leads to
performance enhancement.

We stress that the interaction free measurement device consists 
of absolutely ideal elements. This is certainly not the case 
in praxis and limits the practical value of our model. In case of 
non-ideal detectors and beamsplitters there can be photon absorption 
in the beamsplitter and, consequently, the detectors will not fire. 
In addition, the detector efficiency in real experiment is relatively 
low and the overall efficiency of the conditional quantum eraser 
is strongly reduced. 
On the other side, the conditional quantum eraser is based 
on the detection of all involved photons in an intensity correlation 
experiment. Therefore, whenever photons are lost in case of 
absorption at the beamsplitters or inefficient detectors, these
events will not contribute to the intensity correlation function. 
The no-click event in the measurement device is, ideally, a click
event in the eraser detector. In other words, all involved photon are 
ideally detected, since detectors are placed in both alternatives of 
the beamsplitters. Consequently, inefficient detectors 
simply reduce the overall efficiency of the conditional quantum eraser.
Absorption in the beamsplitters, however, 
changes the ideal partial which way information 
to an effective partial which way information which depends 
on the degree of absorption. This must be taken into account in
realistic experiments and appropriate modifications are necessary. 
These modifications are absorption devices in the other 
optical paths in order to balancing the losses.

\section{Effects of mixture in the initial system}\label{sec.4}
In this section we address the problem if and how some
mixture in the initial two-photon state affects the performance
of the conditional quantum eraser.
We study further its effects on the complementarity relation
between entanglement or concurrence and predictability.
In order to do this we consider the concurrence of a specifically prepared
two-photon system and discuss its connection to the measurable
two-particle visibility.
Suppose we are given the following initial two-photon density operator
\begin{eqnarray}
\rho_{\text{init}}
&=& \frac{1}{1+t}
\Big[t |\pi_{A}\rangle\langle\pi_{A}|\otimes|\sigma_{A}\rangle\langle\sigma_{A}|
\nonumber \\
&& +  |\pi_{B}\rangle\langle\pi_{B}|\otimes|\sigma_{B}\rangle\langle\sigma_{B}|
\nonumber \\
&&
+M\sqrt{t}\left(
|\pi_{A}\rangle\langle\pi_{B}|\otimes|\sigma_{A}\rangle\langle\sigma_{B}|
+ A \leftrightarrow B\right)\Big],
\nonumber \\
&&
\label{eq4.1}
\end{eqnarray}
where $M$ is a ``coherence factor'' bounded between $0\leq C\leq 1$
and can be considered as a measure of mixture in the initial
two-photon state. The two-photon state (\ref{eq4.1}) resembles the
two-photon state (\ref{eq3.5}) of the previous section except for the
reduced coherence expressed by the coherence  factor $M$.
Thus, the biphoton state (\ref{eq4.1}) can be produced by the 
two-atom interferometer in Fig.\ \ref{Fig. 2} and the 
reduced coherence can be generated with the help of the entangled 
atom system. In particular, we assume a decoherence process
on the atoms which become entangled to the $\sigma$-photon
during the generation of the photon system [see Eq.\ (\ref{eq3.3})].
Such a decoherence process can be induced by, e.g.\
a measurement process which determines the population  
in the atomic ground-states, $|1\rangle$ and $|2\rangle$, of one 
of the two atoms \cite{abranyos1}.
Therefore, we may treat the coherence factor $M$ as the effect of a
decoherence process on the atomic system during the generation
of the biphoton in the interferometer, shown in Fig.\ \ref{Fig. 2}. 
The two extremes, $M=1$ and $M=0$, correspond to a pure state 
and a fully incoherent mixed state, respectively. 
 
Let us first investigate the predictability, the two-particle visibility 
and its relation to the concurrence of the given biphoton 
state, (\ref{eq4.1}). These quantities play a crucial role in the 
conventional as well as in the conditional quantum eraser 
as already emphasized in the previous section. 
We are further interested to what extent the degree of mixture 
influences the complementarity relation between concurrence 
and predictability. 
Interestingly, the degree of coherence, $M$, does not have any effect 
on the predictability, ${\cal{P}}$.  The predictability, ${\cal{P}}$, 
is completely determined by the presetting of the interferometer, 
i.e.\ the degree of transmittance $t$ of the optical filter in Fig.\ 
\ref{Fig. 2} (or the transmittance $t_{1}$ of the beamsplitter in 
Fig.\ \ref{Fig. 3} when we employ a partial, interaction-free which-path 
measurement in order to gain knowledge about the photon path),  
\begin{equation}
{\cal{P}} = \frac{1}{1+t}|1-t| = \frac{1-t}{1+t}. 
\label{eq4.2}
\end{equation}
We turn our attention to the two-particle visibility, 
${\cal{V}}_{12}(\pi,\sigma)$, in the initial biphoton system
(\ref{eq4.1}) which is defined as \cite{Jaeger}
\begin{eqnarray}
{\cal{V}}_{12}(\sigma,\pi) &=&
\frac{
[\overline{G}^{\left(2\right)}_{\sigma\pi}(\vec{r},\vec{\rho})]_{\mbox{max}}
-[\overline{G}^{\left(2\right)}_{\sigma\pi}(\vec{r},\vec{\rho})]_{\mbox{min}}}
{[\overline{G}^{\left(2\right)}_{\sigma\pi}(\vec{r},\vec{\rho})]_{\mbox{max}}
+[\overline{G}^{\left(2\right)}_{\sigma\pi}(\vec{r},\vec{\rho})]_{\mbox{min}}}.
\nonumber \\
&& \label{eq4.3}
\end{eqnarray}
Here
\begin{eqnarray}
\overline{G}^{\left(2\right)}_{\sigma\pi}(\vec{r},\vec{\rho})
&=&
{G}^{\left(2\right)}_{\sigma\pi}(\vec{r},\vec{\rho})-
{G}^{\left(1\right)}_{\sigma}(\vec{r}){G}^{\left(1\right)}_{\pi}(\vec{\rho})
\nonumber \\
&&
+|\vec{\Psi}_{\sigma}(\vec{r})|^2|\vec{\Psi}_{\pi}(\vec{\rho})|^2 ,
\label{eq4.4}
\end{eqnarray}
and $G^{\left(2\right)}_{\sigma\pi}=
\langle I_{\sigma}(\vec{r})I_{\pi}(\vec{\rho})\rangle$ 
and $G^{\left(1\right)}_{\sigma,\pi}=\langle I_{\sigma}(\vec{r})\rangle,\,
\langle I_{\pi}(\vec{\rho})\rangle$ are second- and 
first-order intensity correlation functions. The factors 
$|\vec{\Psi}_{\sigma}(\vec{r})|^2$ and $|\vec{\Psi}_{\pi}(\vec{\rho})|^2$ 
express the intensity factors of the dipole radiation, 
\begin{equation}
\vec{E}^{(+)i}_{A,B}(\vec{r},t)=
\Theta\Bigl(t-\frac{|\vec{r}_{A,B}|}{c}\Bigr)
\vec{\Psi}_{i}(\vec{r}){\sigma}^{(-)i}_
{A,B}\Bigl(t-{|\vec{r}_{A,B}|\over c}\Bigr),
\label{eq4.5}
\end{equation}
where
\begin{eqnarray}
\vec{\Psi}_{i}(\vec{r})&=&{-\mu\omega_{o}^2\over 4\pi r^3\epsilon_o c^2}\biggl
[(\vec{\hat\epsilon_i}\times\vec{r})\times\vec{r}\biggr]. 
\label{eq4.6}
\end{eqnarray}
Here, we have introduced the retarded times $t-|\vec{r}_{A,B}|/c$ for
the radiation from atoms $A$ and $B$, respectively. 
Further, $\vec{\hat\epsilon_i}$ forms one of the 
unit vectors $\vec{\hat x}$, $\vec{\hat y}$ or $\vec{\hat z}$, 
depending on the direction of the dipole moment in question, and 
${\sigma}^{(-)i}_{A,B}(t-{|\vec{r}_{A,B}|/ c})$ 
are the ordinary atomic lowering operators for atom $A$ and $B$ 
corresponding to the polarization direction $i$. 
Equation\ (\ref{eq4.4}) contains a correction to the 
intensity correlation $\overline{G}^{\left(2\right)}_{\sigma\pi}$ 
resulting from the fact that 
non-correlated or disentangled photon pairs may also contribute 
to the second-order correlation function 
and this contribution must be subtracted. 
The additional correction, which is proportional to the product of 
constant overall intensities $|\vec{\Psi}_{\sigma}(\vec{r})|^2|
\vec{\Psi}_{\pi}(\vec{\rho})|^2$, must be added back in, 
to compensate for excessive subtraction.  
The excessive subtraction is inherent in the expression 
${G}^{\left(2\right)}_{\sigma\pi}-
{G}^{\left(1\right)}_{\sigma}{G}^{\left(1\right)}_{\pi}$ for entangled 
biphotons \cite{Jaeger}.
The additional correction makes the function (\ref{eq4.3}) 
positive definite or, to be more precise, the two-particle visibility 
lies within the interval $[0,1]$ as it should for a proper definition
of visibility. 

Taking into account the above definitions, we obtain the following 
result for the two-particle visibility of the initial biphoton state 
(\ref{eq4.1}),
\begin{equation}
{\cal{V}}_{12}(\sigma,\pi) = 2M\frac{\sqrt{t}}{1+t}. \label{eq4.7}
\end{equation}
The effect of the coherence factor $M$ is obviously. In the case of a
completely mixed initial 
state, $M=0$, we can no longer observe any two-particle visibility 
even when there is no a priori predictability, ${\cal{P}}=0$, 
which corresponds to $t=1$.  
Let us investigate the relation of two-particle visibility to the 
concurrence, ${\cal{C}}$, in the initial system. 
In principle it is clear, that two-particle visibility must have a
relation to 
entanglement. Two-particle visibility is an explicit 
manifestation of the phase relation between the photons 
in a biphoton system. 
The concurrence of a bipartite, mixed density operator, 
$\rho$, is defined as \cite{wootters} 
\begin{equation}
{\cal{C}}(\rho)={\text{max}}\{0,\lambda_{1}-\lambda_{2}-\lambda_{3} -
\lambda_{4}\}, 
\label{eq4.8}
\end{equation}
where the $\lambda_{i}$'s are the square roots of the 
eigenvalues of $\rho\tilde{\rho}$ in descending order. 
Here $\tilde{\rho}$ results from applying the spin-flip
operation to $\rho^{\ast}$,
\begin{equation}
\tilde{\rho}=(\sigma_{y}\otimes\sigma_{y})\rho^{\ast}
(\sigma_{y}\otimes\sigma_{y}),
\label{eq4.9}
\end{equation}
where $\sigma_{y}$ is the Pauli operator in the relevant standard
basis and $\rho^{\ast}$ is the complex conjugation of $\rho$. 
The relevant basis states in the case of the biphoton, (\ref{eq4.1}),
are $|\pi\rangle_{A}\equiv |0\rangle$ and  
$|\pi\rangle_{B}\equiv |1\rangle$ for the $\pi$-photon system, 
and $|\sigma\rangle_{A}\equiv |0\rangle$ and 
$|\sigma\rangle_{B}\equiv |1\rangle$ for the $\sigma$-photon system. 
With this definition we arrive at the following result for the
concurrence of the initial biphoton state (\ref{eq4.1}),
\begin{equation}
{\cal{C}}(\rho_{\text{init}}) = 
2M \frac{\sqrt{t}}{1+t}, \label{eq4.10}
\end{equation}
which is identical to the measurable two-particle visibility (\ref{eq4.7}). 
We note that a similar relation between concurrence and two-particle 
visibility was found in \cite{sergienko}. 

\begin{figure}[h,t]
\includegraphics[width=8.6cm]{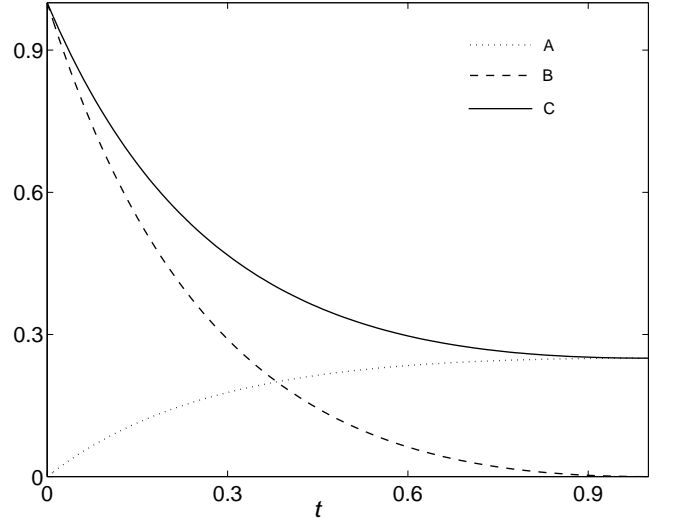}
\caption{Complementarity relation between concurrence or two-particle  
visibility and predictability for a mixed  initial biphoton state 
vs. the transmittance $t$ of the optical filter.  
The coherence factor $M$ is given by  
$M=1/2$ and $A={\cal{C}}^{2}(\rho_{\text{init}})$,  
$B={\cal{P}}^{2}(\rho_{\text{init}})$ and  
$C={\cal{C}}^{2}(\rho_{\text{init}})+{\cal{P}}^{2}(\rho_{\text{init}})$.}
\label{Fig. 4} 
\end{figure}

Let us finally discuss the impact of the mixture on the complementarity 
relation between predictability and concurrence or, 
equivalently, between predictability and two-particle visibility before 
we consider the conditional quantum eraser.  
We already indicated in the previous section that the equal sign in the 
complementarity relation can hold only if the biphoton is in a pure
state. In the case of a mixed state, (\ref{eq4.1}), 
the following complementarity relation 
between concurrence and predictability is found
\begin{eqnarray}
{\cal{C}}^{2}(\rho_{\text{init}})+{\cal{P}}^{2}(\rho_{\text{init}})
&=&
\frac{(1-t)^{2}}{(1+t)^{2}} + 
4M^{2}\frac{t}{(1+t)^{2}}, \nonumber \\
&=&
\frac{(1-t)^{2}}{(1+t)^{2}}+
M^{2}\left[1-\frac{(1-t)^{2}}{(1+t)^{2}}\right], \nonumber \\
&=&
{\cal{P}}^{2}(\rho_{\text{init}})
+M^{2}\left[1-{\cal{P}}^{2}(\rho_{\text{init}})\right]. 
\nonumber \\
&&
\label{eq4.11}
\end{eqnarray}
Thus, the concurrence or equivalently the two-particle visibility 
is bounded by the coherence factor $M$
\begin{equation}
{\cal{C}}(\rho_{\text{init}})\leq M. \label{eq4.12}
\end{equation} 
The concurrence can not reach the optimal value of one even when 
the predictability in the initial biphoton system is equal to zero
(when $t=1$) 
if the degree of coherence or, equivalently, the degree of purity is 
smaller than one $M < 1$. 
We will see that this upper bound on the two-particle visibility or
concurrence  
limits the performance of the conditional quantum eraser. 
The complementarity relation (\ref{eq4.11}) is displayed in Fig.\ 
\ref{Fig. 4} for a decoherence factor of $M=0.5$. The influence of the
decoherence  factor $M$ on the upper bound 
of the complementarity relation as well as on the concurrence is
clearly recognizable. 

We implement the measurement device in the lower left arm of the
interferometer in 
Fig.\ \ref{Fig. 2}, in order to realize the conditional quantum eraser. 
As in the previous section, the biphoton state, which becomes 
coupled to the state of the measurement device, transforms into 
\begin{eqnarray}
&&\!\!\!\!\!\!\!\!\!\!
\rho_{\text{ph}}\otimes\rho_{\text{M}} 
=
\frac{t+t_{\text{BS}}}{1+t}\Big[
\frac{t}{t+t_{\text{BS}}}|\pi_{A}\rangle\langle\pi_{A}|\otimes
|\sigma_{A}\rangle\langle\sigma_{A}|
\nonumber \\
&&+
\frac{t_{\text{BS}}}{t+t_{\text{BS}}}|\pi_{B}\rangle\langle\pi_{B}|\otimes
|\sigma_{B}\rangle\langle\sigma_{B}|
\nonumber \\
&&+
\frac{M\sqrt{t_{\text{BS}}t}}{t+t_{\text{BS}}}
\left(|\pi_{A}\rangle\langle\pi_{B}|\otimes
|\sigma_{A}\rangle\langle\sigma_{B}|   +  A\leftrightarrow B\right)\Big]
\nonumber \\
&&
\otimes |0_{\text{M}}\rangle\langle 0_{\text{M}}| 
\nonumber \\
&&
\frac{1-t_{\text{BS}}}{1+t}
\left(|\pi_{B}^{\left(2\right)}\rangle\langle\pi_{B}^{\left(2\right)}|\otimes
|\sigma_{B}\rangle\langle\sigma_{B}|  
\right)
\otimes |1_{\text{M}}\rangle\langle 1_{\text{M}}| 
\nonumber \\
&& \label{eq4.13}
\end{eqnarray} 
Here, again, $t_{\text{BS}}$ is the transmittance of the beamsplitter 
of the measurement device and $\pi_{B}^{\left(2\right)}$ signifies the 
alternative path of the $|\pi_{B}\rangle$-photon due to the beamsplitter. 
The conditional quantum erasing succeeds if 
the photon detector of the measurement device does not click. 
In this case we realize an interaction-free partial which-way measurement 
of the $\pi$-photon path. 
It is also clear that the $\pi_{B}^{\left(2\right)}$-photon 
is absorbed if the measurement fails, i.e.\ the detector clicks. 
In contrast to the conditional quantum eraser of the previous section, 
which started with a pure entangled biphoton, 
the visibility of the retrieved interference fringes 
under quantum erasing conditions is now bounded by the coherence factor 
$M$. We stress again, that we understand the quantum eraser condition 
as a correlated measurement of the biphoton with an equidistantly 
placed photon detector in between the two atoms for the $\pi$-photon. 
Thus, the visibility we are talking about in the quantum eraser is a 
{\textit{single-particle}} property of the $\sigma$-photon.  
The mixture in the initial state does not have an impact on the success 
or failure probability of the conditional quantum erasing, see 
Eqs.\ (\ref{eq3.7}) and (\ref{eq3.8}). 
However, the maximum possible conditional quantum eraser visibility 
(realized if $t_{\text{BS}}=t$) is bounded by the coherence factor
\begin{equation}
{\cal{V}}^{(\text{QE})}_{\text{cond}}=M
\frac{2\sqrt{tt_{\text{BS}}}}{t+t_{\text{BS}}}
\leq {\cal{V}}^{(\text{QE,max})}_{\text{cond}}
=M. \label{eq4.14}
\end{equation}
Comparing this result with Eq.\ (\ref{eq4.12}) we conclude that 
the concurrence and hence the two-particle visibility limits the 
retrieved conditional quantum eraser visibility. However, we emphasize 
that the retrieved conditional quantum eraser visibility exceeds the 
quantum eraser visibility of the conventional 
quantum eraser ${\cal{V}}^{(\text{QE})}_{\text{max}}$ which is limited by 
\begin{equation}
{\cal{V}}^{(\text{QE})}_{\text{max}}\leq M \sqrt{1-{\cal{P}}^{2}}. 
\label{eq4.15}
\end {equation}
Thus, whenever the a priori knowledge about the path the $\sigma$-photon 
takes is unequal zero, the visibility of the retrieved interference fringes 
in the {\textit{conditional}} quantum eraser tops that of the 
{\textit{conventional}} quantum eraser. In other words the complementarity 
relation between concurrence and predictability which forms 
a physical statement of {\textit{two-particle}} properties 
crucially affects {\textit{single-particle}} features of one 
of the two particles. 

Interestingly, the {\textit{product}} of the success probability 
and the visibility of the recovered interference 
fringes in the {\textit{conditional}} quantum eraser 
is bounded by Eq.\ (\ref{eq4.15}) as the visibility by itself is 
in the conventional quantum eraser. 
This clearly demonstrates how the conditional quantum eraser operates. 
The conditional quantum eraser works successfully if and only if 
the detector of the partial which-way measurement device does not 
click. In other words, a non-unitary projection onto a certain subspace 
takes place. 
We mention finally, that the above consideration can easily be transferred 
to the interference device in Fig.\ \ref{Fig. 3} similar as in the previous 
section. The only difference is the which-path 
knowledge gained from the additional measurement device for 
the $\sigma$-photon instead of the a priori which-path knowledge. 
\section{Conclusions}\label{sec.5}
The usual quantum eraser problem considers the erasure of 
{\textit{possible}} but not actual which-way information. 
Here, we have considered a different task. We study a 
{\textit{conditional}} quantum eraser to erase a priori or 
{\textit{actual}} which-path information. 
The conditional quantum eraser employs a partial, interaction-free 
measurement on the erasing photon which crucially influences on 
the interfering photon if the two photons form an entangled 
state. The conditional quantum eraser erases 
the a priori knowledge of the interfering photon completely. 
However, this can only be achieved with a certain success probability 
which explains the conditional operation.  
Thus, the conditional quantum eraser constitutes a non-unitary
transformation on the biphoton. 
The visibility of the retrieved interference fringes exceeds that of 
a conventional quantum eraser which is bounded according to 
a complementarity relation between concurrence or two-particle 
visibility and predictability. 
In particular, the visibility of the recovered interference fringes 
in a {\textit{conventional}} quantum eraser can not exceed a certain 
limit set by the a priori knowledge about the path the interference photon 
takes. In other words, a conventional quantum eraser can not 
erase predictability or actual which-path information. 
On the other hand, the complementarity relation between 
the concurrence and predictability limits the performance 
of the {\textit{conditional}} quantum eraser, as well. 
Especially in cases when the initial biphoton system contains 
some degree of mixture, the visibility of the retrieved interference 
fringes can not exceed the upper bound of the inequality set by 
the complementarity relation. 
This generates an interesting scenario where two-particle properties 
crucially influence single-particle properties. 
The complementarity relation between concurrence and predictability 
opens an interesting area of investigations which focuses on connections 
between complementarity and nonlocality. In this context we note that a
similar complementarity relation was found in 
\cite{bose}. However, the relation between concurrence and the observable 
two-particle visibility as well as the influences of mixture in the 
biparticle was not investigated there. 

In conclusion, we have studied the conditional quantum eraser in a
two-atom, four-port interferometer. 
The two atoms substitute a Young-type double-slit an generate an 
entangled photon pair. 
The two photons are orthogonally polarized and thus distinguishable. 
One of the photons is detected at an interference screen
(the interference photon) while the other one is detected 
at an equidistant position from the two atoms (the erasing photon). 
This establishes the conventional quantum eraser scenario. 
The conditional quantum eraser additionally performs a partial and 
interaction-free which-path measurement on the erasing photon. 
The interaction-free, partial which-path measurement can be 
simulated with a beamsplitter and an additional detector.  
The detector indicates whether the conditional quantum eraser has 
succeeded or failed. 
In particular, a detector click results in a failure of the 
conditional quantum eraser. 
The entangled atom-photon system, during the generation of the biphoton, 
enables us to study effects of mixedness in the biphoton state on the
performance of the conditional quantum eraser. In addition, effects on 
the complementarity relation between two-particle visibility and 
predictability can be explored. A measurement process on the atoms 
which introduces decoherence effects \cite{abranyos1} may be employed 
in order to generate a certain amount of mixture in the biphoton system. 
Finally, we note that the Young-type interference experiment 
was realized by Eichmann {\textit{et al.}} \cite{eichmann1} 
and the implementation of a conditional quantum eraser 
should be experimentally feasible. 
\begin{acknowledgments}
We acknowledge helpful discussions with Y.\ Abranyos, 
M.\ Hillery, I.\ Nemeth, S.\ Stenholm, and Y.\ Sun. 
This research was supported by a grant from the Office of Naval 
Research (grant No. N00014-92J-1233), by the European
Union Research and Training Network COCOMO, Contract No.\
HPRN-CT-1999-00129 and by a grant from PSC-CUNY.
\end{acknowledgments}

\begin{thebibliography}{99}
\bibitem{scully1} M.\ O.\ Scully and K.\ Dr\"uhl, \pra {\bf 25}, 2208 (1982).
\bibitem{bohr} N.\ Bohr, Naturwissenschaften {\bf 16}, 245 (1928);
Nature (London) {\bf 121}, 580 (1928).
\bibitem{scully2} M.\ Hillery and M.\ O.\ Scully,
{\it{Quantum Optics, Experimental Gravity, and Measurement Theory}}
(Plenum, NY, 1983) p.\ 65.
\bibitem{englert1} M.\ O.\ Scully, B.\ G.\ Englert, and H.\ Walther,
Nature (London) {\bf 351}, 111 (1991).
\bibitem{abranyos1} Y.\ Abranyos, M.\ Jakob, and J.\ Bergou,
\pra {\bf 60} R2618 (1999); Acta Phys.\ Slov.\ {\bf 48}, 255 (1998). 
\bibitem{jakob} M.\ Jakob, Y.\ Abranyos, and J.\ A.\ Bergou, 
\pra {\bf 64}, 062102 (2001). 
\bibitem{zajonc}
A.\ G.\ Zajonc, L.\ J.\ Wang, X.\ Y.\ Zou, and L.\ Mandel, Nature (London)
$\bf{353}$, 507 (1991).
\bibitem{kwiat}
P.\ G.\ Kwiat, A.\ M.\ Steinberg, and R.\ Y.\ Chiao, 
\pra {\bf 11}, 7729 (1992).
\bibitem{zeilinger1}
Th.\ J.\ Herzog, P.\ G.\ Kwiat, H.\ Weinfurter, and A.\ Zeilinger,
\prl {\bf 75}, 3034 (1995).
\bibitem{Pittman} T.\ B.\ Pittman, D.\ V.\ Strekalov, A.\ Migdall,
M.\ H.\ Rubin, A.\ V.\ Sergienko, and Y.\ H.\ Shih, 
\prl {\bf 77}, 1917 (1996).
\bibitem{Duerr} S.\ D\"urr, T.\ Nonn, and G.\ Rempe, Nature (London)
{\bf 395}, 33 (1998).
\bibitem{Kim}Y.-H.\ Kim R.\ Yu, S.\ P.\ Kulik, Y.\ Shih, and M.\ O.\ Scully,
\prl {\bf 84}, 1 (2000).
\bibitem{Zurek} W.\ K.\ Wooters and W.\ H.\ Zurek, \prd {\bf 19}, 473
(1979).
\bibitem{Greenberger}
D.\ M.\ Greenberger and A.\ Yasin, Phys.\ Lett.\ A {\bf 128}, 391
(1988).
\bibitem{Mandel}
L.\ Mandel, Opt.\ Lett.\ {\bf 16} 1882 (1991).
\bibitem{Rauch}
H.\ Rauch and J.\ Summhammer, Phys.\ Lett.\ A {\bf 104}, 44 (1984);
J.\ Summhammer, H.\ Rauch, and D.\ Tuppinger, \pra {\bf 36}, 4447
(1987).
\bibitem{englert} B.\ G.\ Englert, \prl {\bf 77}, 2154 (1996).
\bibitem{Jaeger}
G.\ Jaeger, A.\ Shimony, and L.\ Vaidman, \pra {\bf 51}, 54 (1995).
\bibitem{Duerr1} S.\ D\"urr, T.\ Nonn, G.\ Rempe, \prl {\bf 81}, 5705 (1998).
\bibitem{Tsegaye} T.\ Tsegaye, G.\ Bj\"ork, A.\ Atat\"ure, A.\ V.\ Sergienko,
B.\ E.\ A.\ Saleh, And M.\ C.\ Teich, \pra {\bf 62}, 032106 (2000).
\bibitem{bjoerk} G.\ Bj\"ork and A.\ Karlsson, \pra {\bf 58}, 3477 (1998).
\bibitem{abranyos2} Y.\ Abranyos, M.\ Jakob, and J.\ Bergou,
\pra {\bf 61}, 013804 (2000).
\bibitem{englert2} B.-G.\ Englert and J.\ A.\ Bergou, 
\oc {\bf 179}, 337 (2000).
\bibitem{Elitzur}
A.\ C.\ Elitzur and S.\ Dolev, \pra {\bf 63}, 062109 (2001).
\bibitem{schuurman}
D.\ Polder and M.\ F.\ H.\ Schuurmans, \pra {\bf 14}, 1468 (1976).
\bibitem{eichmann1}
U.\ Eichmann, J.\ C.\ Bergquist, J.\ J.\ Bollinger, J.\ M.\ Gilligan, 
W.\ M.\ Itano, D.\ J.\ Wineland, and M.\ G.\ Raizen,
\prl {\bf 70}, 2359 (1993);
W.\ M.\ Itano, J.\ C.\ Bergquist, J.\ J.\ Bollinger, D.\ J.\ Wineland,
U.\ Eichmann, and M.\ G.\ Raizen, \pra {\bf 57}, 4176 (1998).
\bibitem{interference} see for example: P.\ Kochan, H.\ J.\ Carmichael,
P.\ R.\ Morrow, and M.\ G.\ Raizen, \prl {\bf 75}, 45 (1995);
R.\ G.\ Brewer, {\textit{ibid.}} {\bf 77}, 5153 (1996);
T.\ Wong, S.\ M.\ Tan, M.\ J.\ Collett, and D.\ F.\ Walls,
\pra {\bf 55}, 1288 (1997);
G.\ M.\ Meyer and G.\ Yeoman, \prl {\bf 79}, 2650 (1997);
P.\ D.\ D.\ Schwindt, P.\ G.\ Kwiat, and B.-G.\ Englert, 
\pra {\bf 60}, 4285 (1999);
H.\ T.\ Dung and K.\ Ujihara, \prl {\bf 84}, 254 (2000);
A.\ Luis, \pra {\bf 64}, 012103 (2001);
C.\ Sch\"on and A.\ Beige, {\it ibid.} {\bf 64}, 023806 (2001);
C.\ Skornia, J.\ von Zanthier, G.\ S.\ Agarwal, E.\ Werner, and H.\ Walther,
{\it ibid.} {\bf 64}, 063801 (2001); 
G.\ S.\ Agarwal, J.\ von Zanthier, C.\ Skornia, and H.\ Walther, 
{\it ibid.} {\bf 65}, 053826 (2002).
\bibitem{wootters}
see for example: W.\ K.\ Wootters, Quantum Inf.\ Comp.\ {\bf 1}, 27 (2001).
\bibitem{sergienko}
A.\ F.\ Abouraddy, B.\ E.\ A.\ Saleh, A.\ V.\ Sergienko, and M.\ C.\ Teich,
\pra {\bf 64}, 050101(R) (2001). 
\bibitem{bose} S.\ Bose and D.\ Home, \prl {\bf 88}, 050401, (2002).
\end{thebibliography}
\end{document}